\documentclass[aps,prl,twocolumn,superscriptaddress]{revtex4-1}
\usepackage{graphicx,amsmath}
\usepackage{color}

\begin{document}

%\preprint{}

\title{A Differential Approximation Model For Passive Scalar Turbulence}


\author{P. Morel}
\author{Shaokang Xu}
\author{\"O.~D.~G\"urcan}
%\thanks{}
\affiliation{LPP, Universit\'e Paris-sud, \'Ecole Polytechnique,
  CNRS, Sorbonne Universit\'e, 91128 Palaiseau}

%\date{\today}

\begin{abstract}
  Two dimensional passive scalar turbulence is studied by means of a k-space diffusion model based on a third order differential approximation. This simple description of local nonlinear interactions in Fourier space is shown to present a general expression, in line with previous seminal works, and appears to be suitable for various 2D turbulence problems. Steady state solutions for the spectral energy density of the flow is shown to recover the Kraichnan-Kolmogorov phenomenology of the dual cascade, while various passive scalar spectra, such as Batchelor or Obukhov-Corssin spectra are recovered as steady state solutions of the spectral energy density of the passive scalar. These analytical results are then corroborated by numerical solutions of the time evolving problem with energy and passive scalar injection and dissipation on a logarithmic wavenumber space grid over a large range of scales. The particular power law spectra that is obtained is found to depend mainly on  the location of the kinetic and passive scalar energy injections.
%  Turbulent spectra are studied analytically as well as numerically in the context of strongly magnetized plasmas. Diffusive forms in $k$-space are derived as {\it continuum} limit of isotropic turbulence, where the nonlinear interactions are considered strongly local in $k$-space, or alternatively mediated by large scales, {\it ie} associated to disparate scales interactions. Different spectral functionnal are found, depending on the form of the quasi-neutrality equation.
\end{abstract}

%\pacs{}
\maketitle

Differential approximation models, also referred to as nonlinear diffusion models, are simple cascade models that rely on the assumptions of isotropy and locality of energy transfers in $k$-space. While generally questionable, such assumptions can sometimes be justified and are linked to the pioneering works on turbulence \cite{Kolmogorov_CRAR_1941, Frisch_Turbulence_Book}. In a seminal study, Kovalsznay \cite{Kovalsznay_JAeroSci_1948} proposed a first order diffusion term to model the energy transfers in a turbulent {\it medium}, allowing to recover the $k^{-5/3}$ energy power law, universal for 3D turbulence \cite{Kolmogorov_CRAR_1941}. Inspired by the ability of such models to describe also radiation and neutron transport \cite{Davidson_1958}, Leith extended such a model to second order derivative, allowing to recover the spectral corrections due to the transition from inertial to viscous spectral domain \cite{Leith_PoF_1967}. Such models have also been applied to the case of 2D turbulence \cite{Leith_PoF_1968, Lilly_JAS_1989}, where the dual cascade results in the well known Kraichnan-Kolmogorov {\it spectra}, combining $k^{-5/3}$ and $k^{-3}$ power laws \cite{Kraichnan_PoF_1967, Batchelor_PoF_1969, Boffetta_ARFM_2012}.

By using a fourth order derivative, nonlinear diffusion models have also been extended in order to capture the thermodynamic {\it equilibrium spectra} in addition to the two cascade {\it spectra} in 2D turbulence \cite{Lvov_JETP_2006}. Extensions of such models to magnetohydrodynamics and gyrofluids have been proposed and progressively enriched \cite{Iroshnikov_1964, Zhou_1990, Lithwick_2003, Matthaeus_2009, Galtier_ApJ_2010, Passot_JPP_2019}. Interestingly, Thalabard {\it et al} generalized, by means of a generic family of differential approximation models \cite{Thalabard_JPA_2015}, the possible existence of anomalous spectral laws during the transient development of the inertial range {\it spectrum} showed first by Connaughton using a Leith' model \cite{Connaughton_PRL_2004} called ``warm cascades''. Such transient behavior could explain the observed slope of the solar wind in the vicinity of the Earth \cite{David_ApJ_2019}.

Notably, such models have also been applied to superfluid turbulence \cite{Lvov_JLTP_2008, Nazarenko_JETP_2007, Nazarenko_JETPPL_2006}, wave turbulence in a large variety of physical systems \cite{Zakharov_NPG_1999, Nazarenko_JSM_2006, Pushkarev_PhysD_2000, Boue_PRB_2011, Galtier_JPP_2000}, and very recently to gravitational wave turbulence \cite{Galtier_PhysD_2019}.

Since the form of its nonlinear term is generic to a wide class of 2D turbulence problems, it makes sense to base our analysis on the problem of two dimensional passive scalar turbulence, which couple the dynamical evolution of the vorticity $\nabla^2 \phi$ field, associated to a flow field ${\bf u} = - {\bf z} \times \nabla \phi$, to the evolution of a passive scalar field $n$, simply advected by this flow:
\begin{eqnarray}
 \label{eq:vorticity}
 \partial_t \nabla^2 \phi + {\bf z} \times \nabla \phi \, . \, \nabla \nabla^2 \phi & = & I_\phi - \nu \triangle \nabla^2 \phi, , \\
 \label{eq:passive}
 \partial_t n + {\bf z} \times \nabla \phi \, . \, \nabla n & = & I_n - \kappa \triangle n \, .
\end{eqnarray}

Such equations cover a wide variety of physical systems, since the observable $n$ can refer to the fluid density, to the fluid temperature (if assumed isotropic and if the buoyancy force is neglected), but it can also represent the density of some passive impurities, or, in the case of magnetized plasma turbulence, a charge density advected by a turbulent plasma ${\bf E} \times {\bf B}$ flow \cite{Hasegawa_PRL_1983}.

From an energetic point of view, and assuming locality of the interactions in $k$-space, the equations governing the two unknowns (\ref{eq:vorticity}, \ref{eq:passive}), concern respectively the kinetic energy $\mathcal{E}_\phi = k^2 \left \| \phi_k \right \|^2$ and the passive scalar energy $\mathcal{E}_n  = \left \| n_k \right \|^2$, and can be modeled as follows:
\begin{eqnarray}
 \partial_t \mathcal{E}_\phi + \partial_k \Pi_k^{\mathcal{E}_\phi} & = & \mathcal{I}_k^\phi - D_\phi \mathcal{E}_\phi \, , \label{eq:dEphidt} \\
 \partial_t \mathcal{E}_n + \partial_k \Pi_k^{\mathcal{E}_n} & = & \mathcal{I}_k^n - D_n \mathcal{E}_n \, , \label{eq:dEndt}
\end{eqnarray} where explicit forms of the energy fluxes $\Pi_k^{\mathcal{E}_\phi}$ and $\Pi_k^{\mathcal{E}_n}$ remains undefined. The injections $\mathcal{I}_k^\phi$, $\mathcal{I}_k^n$ will be assumed constant with gaussian shape along $k$: $\mathcal{I}_k^{\phi, n} = e^{- (x-x_0^{\phi, n})/(2 \Delta x^2)}$, where $x$ is such that $k = k_{\textnormal {\tiny min}} e^{x}$, and so are the injection scales $k_0^{\phi, n} = k_{\textnormal{\tiny min}} \, e^{x_0^{\phi, n}}$. Dissipation is ensured by diffusion operators acting mainly at small scales (with a viscosity $\nu_s^{\phi, n}$), and hypodiffusion operators acting at largest scales (with $\nu_L^{\phi, n}$), in order to prevent the inverse cascade leading to an accumulation of energy at the largest scales $D_{n, \phi} = \nu_L^{\phi, n} k^{-6} + \nu_s^{\phi, n} k^2 \, .$

\paragraph{The model for fluxes:-}

Differential approximation models offer a simple reformulation of the nonlinear term, assuming locality of the nonlinear couplings in $k$-space. Since the Euler equation (\ref{eq:vorticity}) formally has the same form as that of the passive scalar one (\ref{eq:passive}) but with $n$ replaced by $\nabla^2 \phi$, here we focus on the spectral energy flux of the passive scalar. In order to propose a particular form for it, we first note that since the nonlinearity in the equation (\ref{eq:passive}) for $n$ has the form of a Poisson bracket, divergence of the flux which represents this nonlinear term (multiplied by $n_k/k$), should reflect this. In particular, recall that the Poisson bracket changes sign when $n$ and $\phi$ are exchanged, and that it vanishes when $\phi$=$n$. This basically means that a term of the form $\partial_k \frac{\phi}{n} = \partial_k \frac{\sqrt{\mathcal{E}_\phi}}{k \sqrt{\mathcal{E}_n}}$ should appear as one of the inner derivatives. Moreover, since the nonlinear term in the passive scalar energy budget is linear with respect to $\phi_k$, this should be the only place where $\mathcal{E}_\phi$ appears in the flux. These observations lead to the following general form for the spectral energy flux of the passive scalar:
\begin{equation}
 \label{eq:flux_general}
 \Pi_k^{\mathcal{E}_n} = C \left [ k^\alpha \mathcal{E}_n^\beta \partial_k \left ( k^{\alpha'} \mathcal{E}_n^{\beta'} \partial_k \frac{\sqrt{\mathcal{E}_\phi}}{k \sqrt{\mathcal{E}_n}} \right ) \right ] \, .
\end{equation}

Dimensional analysis applied to the first two terms of the passive scalar spectral energy budget (\ref{eq:dEndt}) with this flux (\ref{eq:flux_general}), gives us:
\begin{equation}
 \left [ \mathcal{E}_n \right ] T^{-1} = L^{3 - \alpha - \alpha' + 3/2 + 1} T^{-1} \left [ \mathcal{E}_n \right ]^{\beta + \beta' - 1/2} \, ,
 \label{eq:dimensionnal}
\end{equation} where the dimension of the spectral kinetic energy density $\left [ \mathcal{E}_\phi \right ] = L^3 T^{-2}$ has been used. According to equation (\ref{eq:passive}), the previous relation should not depend on the dimension of the passive scalar energy $\mathcal{E}_n$. This fixes $\alpha + \alpha' = 11/2$, together with $\beta + \beta' = 3/2$.

Furthermore, noting that if we let $\mathcal{E}_n = k^2 \mathcal{E}_\phi$, we should recover the enstrophy flux, as represented for instance in the 2D Leith model [i.e. $\displaystyle \Pi_k^{\mathcal{W}_\phi}=-k^2\Pi_k^{\mathcal{E}_\phi}$ where
$\displaystyle \Pi_k^{\mathcal{E}_\phi} = - C k^{-1} \partial_k \left ( k^{9/2} \mathcal{E}_\phi^{3/2} \right )$], gives $\alpha' = 11/4$, and $\beta' = 3/4$, and the model for the nonlinear fluxes finally reads as follows:
\begin{eqnarray}
 \label{eq:Pikn}
 \Pi_k^{\mathcal{E}_n} & = & C \left [ k^{11/4} \mathcal{E}_n^{3/4} \partial_k \left ( k^{11/4} \mathcal{E}_n^{3/4} \partial_k \frac{\mathcal{E}_\phi^{1/2}}{k \mathcal{E}_n^{1/2}} \right ) \right ] \, , \\
 \label{eq:Pikphi}
 \Pi_k^{\mathcal{E}_\phi} & = & - C \frac{\partial_k \left ( k^{9/2} \mathcal{E}_\phi^{3/2} \right )}{k} \, .
\end{eqnarray} 

These expressions hold for a variety of 2D physical systems, where a physical quantity $f$ is advected by an incompressible 2D flow ${\bf u} = {\bf z} \times \nabla \phi$, resulting in a Poisson bracket structure of the nonlinear advection term $\propto {\bf u} . \nabla f$.

Away from the injection and dissipative ranges, the quasi-stationarity of the kinetic energy flux (\ref{eq:Pikphi}) gives the two scalings $\mathcal{E}_\phi (k) \sim k^{-3}$, $\mathcal{E}_{\phi} (k) \sim k^{-5/3}$, corresponding to the Kraichnan-Kolmogorov {\it spectrum}, characteristic of 2D turbulence \cite{Frisch_Turbulence_Book, Batchelor_PoF_1969, Kraichnan_PoF_1967}. As summarized in Table \ref{table:spectrae}, the spectral domain associated to the $-5/3$ slope corresponds to an inverse energy cascade with zero enstrophy flux, while the domain with the $-3$ power law experiences zero energy flux with a forward enstrophy cascade.

By inserting these two solutions, the stationarity condition: $\partial_k \Pi_k^{\mathcal{E}_n} = 0$, can be solved with respect to $\mathcal{E}_n$:
$$\mathcal{E}_n (k) \sim \left \{ k^{-5} \, ; \, k^{-11/3} \, ; \, k^{-5/3} \, ; \, k^{-1} \, ; \, k^{1/3} \, ; \, k^3 \right \} \, .$$

Among those solutions, the Corrsin-Obukhov \cite{Corrsin_JAppPhys_1951, Obukhov_IzvAkadNauk_1949} and the Batchelor \cite{Batchelor_JFM_1959} {\it spectra}, are associated with a positive passive scalar energy flux, and with the slopes $-5/3$ and $-1$, respectively. One should mention that various other power laws have been discussed in the literature, mainly by varying the Prandtl number $Pr \equiv \nu_s^\phi / \nu_s^n$.
In the small Prandtl regime, EDQNM approach predicts a steepening of the $k^{-5/3}$ slope into $k^{-17/3}$ and $k^{-11/3}$ at smallest scales \cite{Briard_PRE_2015}. The case $k^{1/3}$ has also been observed (together with $k^{1}$ and $k^{-1}$), with help of a renormalization group analysis of the large scales behavior \cite{Mazzino_JSM_2009}, as well as by means of EDQNM \cite{Lesieur_JFM_1985}.

\begin{center}
\begin{table}
 \begin{tabular}{c||c|c|c|c|c|c}
  %\hline \hline
%   $\mathcal{E}_\phi$ & \, $\propto k^{-5/3}$ \, & \, $\propto k^{-5/3}$ \, & \, $\propto k^{-5/3}$ \, & \, $\propto k^{-3}$ \, & \, $\propto k^{-3}$ \, & \, $\propto k^{-3}$ \,  \\ \hline 
  $\mathcal{E}_\phi$ & \multicolumn{3}{c|}{$\propto k^{-5/3}$} & \multicolumn{3}{|c}{$\propto k^{-3}$}  \\ \hline 
  $\mathcal{E}_n$ & \, $\propto k^{-11/3}$ & \, $\propto k^{-5/3}$ & \, $\propto k^{1/3}$ & \, $\propto k^{-5}$ & \, $\propto k^{-1}$ & \, $\propto k^{3}$ \\ \hline
  $\Pi_k^{\mathcal{E}_\phi}$ & $<0$ & $<0$ & $<0$ & $0$ & $0$ & $0$ \\ \hline
  $\Pi_k^{\mathcal{E}_n}$ & $0$ & $>0$ & $0$ & $0$ & $>0$ & $0$ %\\ \hline \hline
 \end{tabular}
 \caption{Summary of the scalings allowed for the energies $\mathcal{E}_\phi$ and $\mathcal{E}_n$, and sign of the associated constant fluxes.}
 \label{table:spectrae}
\end{table}
\end{center}

\paragraph{Numerical integration:-}

The numerical integration of the equations (\ref{eq:dEphidt}, \ref{eq:dEndt}) is performed with a fourth order adaptive Runge Kutta solver from scipy python library\cite{Travis_CompSciEng_2007}, while the $k$ derivatives are evaluated by means of centered second order finite differences on a logarithmic grid $k_i = k_{\textnormal {\tiny min}} e^{x_i}$. Such a logarithmic grid allows to cover a large range of decades at a very modest numerical cost.

\begin{figure}
 \includegraphics[width=\linewidth]{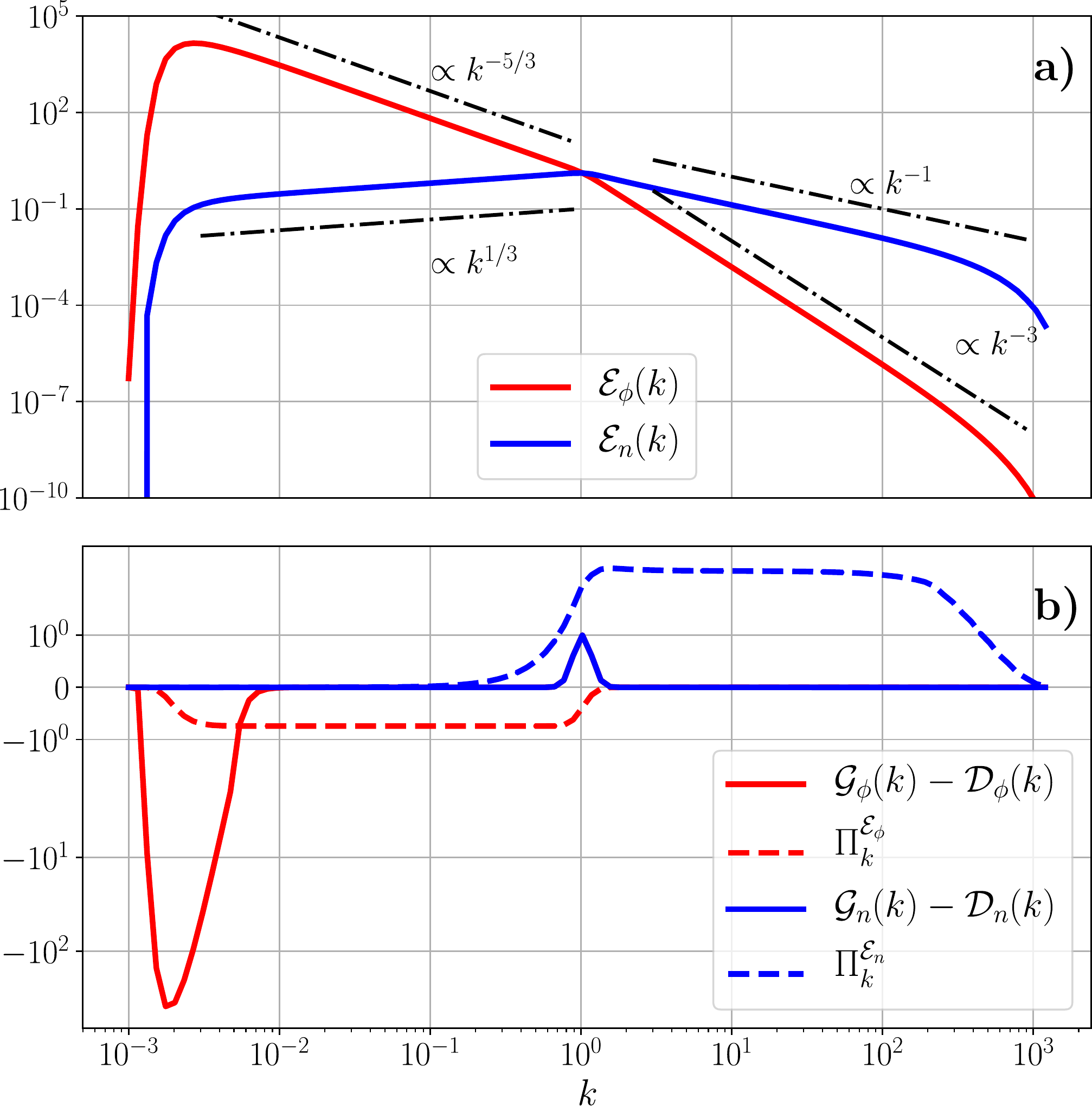}
 \caption{Kinetic an passive scalar energy spectra (a), associated injections, fluxes and dissipations (b with symmetric logarithmic $y$-scale). The injections of kinetic energy and passive scalar energy are located at $k_0 = 1$. A symmetrized logarithmic (symlog) scale is used in the bottom part of the figure.}
 \label{fig:Ep_En_k}
\end{figure}

The kinetic $\mathcal{E}_\phi (k)$, and passive scalar energies $\mathcal{E}_n (k)$ are shown in figure \ref{fig:Ep_En_k} (a), as functions of the wave vector $k$. The injections are located at $k_0 = 1$, for both energies. In figure \ref{fig:Ep_En_k} (b), the sum of the injection and dissipations (solid line), as well as the associated flux (dashed line), are given for both the kinetic energy (in red), and the passive scalar energy (in blue).

Regarding the kinetic energy, the usual Kraichnan-Kolmogorov spectra are recovered numerically, where the $k^{-5/3}$ slope is found at scales larger than the injection scale, associated with the inverse cascade of energy $\Pi_k^{\mathcal{E}_\phi} < 0$. The scales smaller than the injection scale exhibit a $k^{-3}$ spectrum, corresponding to a direct cascade of enstrophy. On the other hand, the passive scalar energy $\mathcal{E}_n (k)$ displays a Batchelor spectrum $k^{-1}$ at small scales, corresponding to a forward cascade $\Pi_k^{\mathcal{E}_n} > 0$, while a $k^{1/3}$ slope is found at the largest scales, corresponding to a zero energy flux solution $\Pi_k^{\mathcal{E}_n} = 0$. 

\begin{figure}
 \includegraphics[width=\linewidth]{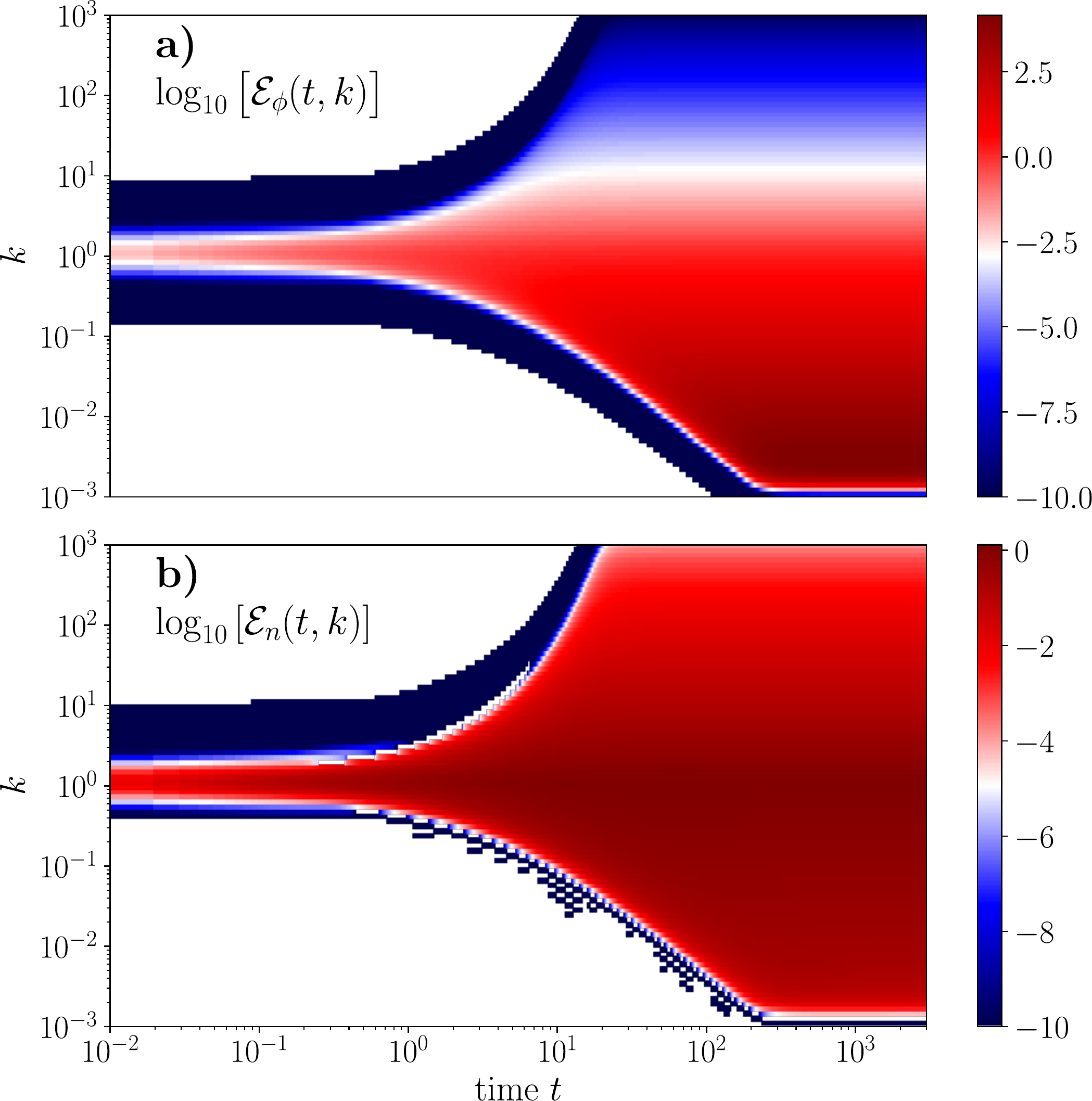}
  \caption{Temporal evolution of the kinetic energy {\it spectrum} (a), and passive scalar energy (b). Both $k$ and time are given in log-scale. At early times, the {\it spectra} are not filling the whole $k$-range, leading to the two white areas obtained on each figure (not related to the color bar).}
 \label{fig:Epkt}
\end{figure}

In figure \ref{fig:Epkt}, the kinetic (a), and the passive scalar energy (b), are given as functions of both time and $k$, in log-log scales. In both cases, the {\it spectrum} exhibits a smooth time evolution, thanks to the energy formulation of the equations that eliminates phase dynamics. Thus, the differences between instantaneous or time averaged {\it spectra} are insignificant. Notably the {\it spectra} converge faster toward small scales than towards the large scales, by about one order of magnitude.

\begin{figure}
 \includegraphics[width=\linewidth]{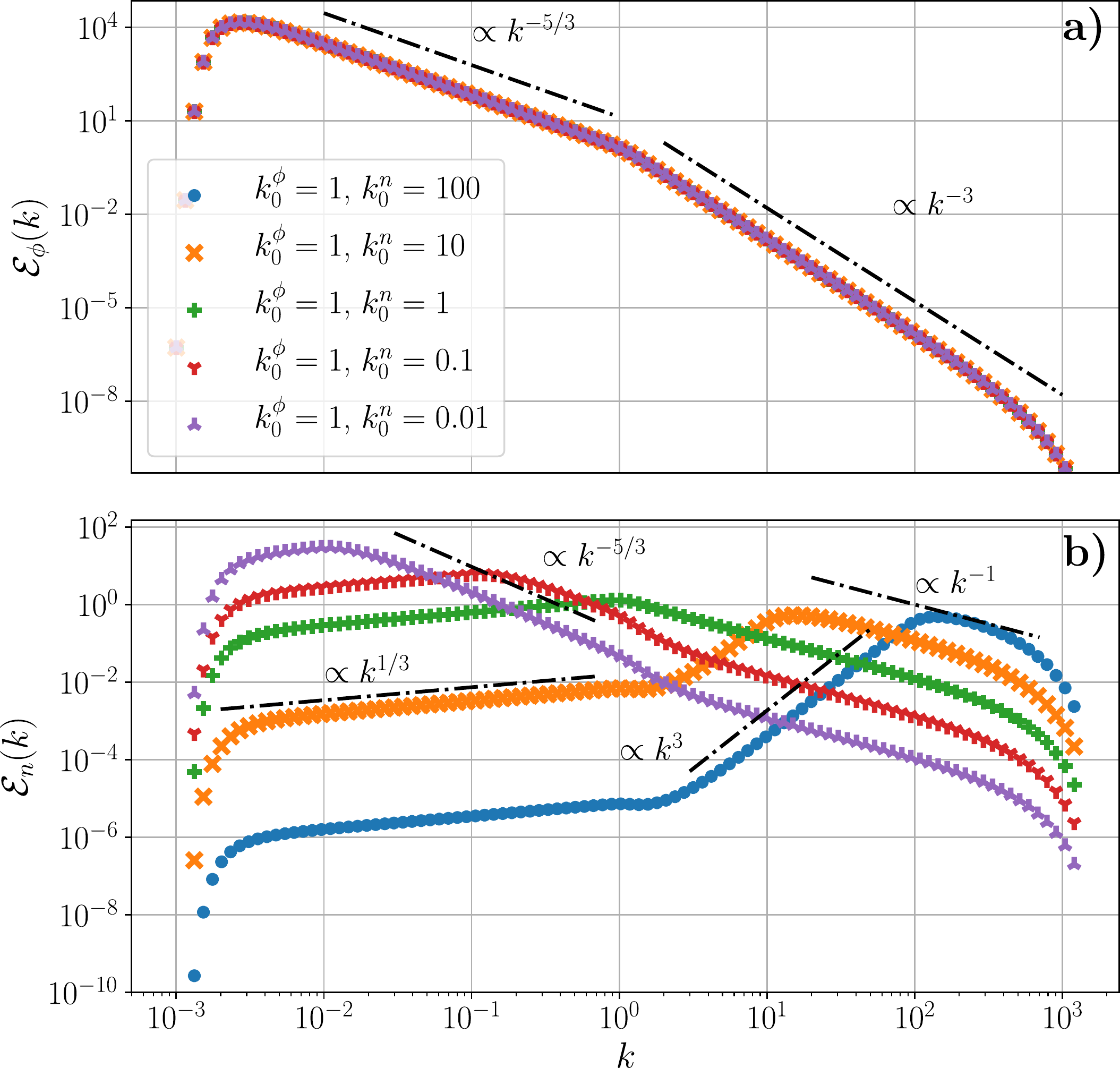}
 \caption{Kinetic energy $\mathcal{E}_\phi$ (a) and passive scalar energy $\mathcal{E}_n$ (b) as functions of the wave vector $k$, for various locations of the passive scalar injection $k_0^n = 100, 10, 1, 0.1, 0.01$. The injection of kinetic energy is located at $k_0^\phi = 1$.}
 \label{fig:Ephi_En_k_scan_Ikn}
\end{figure}

The results of a second study, investigating the role of the passive scalar energy injection scale $\mathcal{I}_k^n = e^{-(x - x_0^n)^2 / (2 \Delta x^2)}$ on the resulting wave-number spectra of passive scalar energy density are shown in figure \ref{fig:Ephi_En_k_scan_Ikn}. The reference case is the same as discussed previously.

Passive scalar energy {\it spectrum} is strongly affected by the energy injection scale, as follows. First, if the passive scalar injection scale is larger than the kinetic energy one ($k_0^n < k_0^\phi$), a $k^{-5/3}$ slope appears in order to connect the $k^{1/3}$ range at largest scales to the $k^{-1}$ at smallest scales. In other words, when an inverse kinetic energy cascade coexists with a forward passive scalar cascade, both the kinetic energy and the passive scalar energy exhibits a $k^{-5/3}$ slope, characteristic of the Corrsin-Obukhov spectrum \cite{Obukhov_IzvAkadNauk_1949, Corrsin_JAppPhys_1951}. Second, if the passive scalar energy is injected at smaller scales than that of kinetic energy ($k_0^n > k_0^\phi$), a $k^3$ slope is observed for the passive scalar energy to make the link between the range where $\mathcal{E}_n \sim k^{1/3}$ at small scales associated to the inverse energy cascade, and the one where the Batchelor's {\it spectrum} $\mathcal{E}_\phi \sim k^{-1}$ appears, at smallest scales combining forward cascades of kinetic and passive scalar energies. Such a $k^3$ power law is associated to a zero passive scalar energy flux, combined with a direct cascade of kinetic energy.

\paragraph{Discussion:-}

A diffusive form representing local energy transfers in $k$-space is proposed to model the effect of nonlinear interactions in the case of 2D isotropic passive scalar turbulence. Such a description is shown to reproduce well known existing results such as the Batchelor {\it spectrum} $\mathcal{E}_n \propto k^{-1}$ corresponding to a forward cascade of passive scalar energy density in a forward enstrophy cascade range $\mathcal{E}_\phi \propto k^{-3}$. In contrast, in the inverse energy cascade $\mathcal{E}_\phi \propto k^{-5/3}$, if the passive scalar is injected at larger scales, one obtains a forward cascade for passive scalar energy with the scaling $\mathcal{E}_n \propto k^{-5/3}$, which is the Corrsin-Obukhov {\it spectrum}. On the other hand, if the passive scalar is injected at the same scale as the energy, then the inverse kinetic energy cascade is accompanied by a positive passive scalar energy slope $\mathcal{E}_n \propto k^{1/3}$, with zero flux.

One could question the locality assumption, especially considering Batchelor's theory is based on the interactions between small scale passive scalar eddies with large scale flow structures. But since $\mathcal{E}_n \propto k^{-1}$ is accompanied by $\mathcal{E}_\phi \propto k^{-3}$, the kinetic energy content is far higher at large scales, which indeed corresponds to a case of large scale vortices coexisting with small scale passive scalar turbulence, even though the interactions may remain local. In that sense, such a result comes in line with the idea that the Batchelor's {\it spectrum} is more resilient than its underlying hypothesis \cite{Kraichnan_PoF_1968}.

The diffusion approximation given here (\ref{eq:Pikn}) may be used in many physical problems involving a nonlinear advection term under a Poisson bracket form implied by two dimensional turbulence, such as magnetized plasmas, geophysical fluids, or certain laboratory experiments to count a few examples. Note that, the main effort in modeling was that of the passive scalar energy flux and the modeling of the nonlinear flux of kinetic energy comes as a corollary.

\paragraph{Acknowledgements:-}
The authors would like to thank P.~H.~Diamond, T.~S.~Hahm, J.~Anderson, and the participants of the Festival de Th\'eorie for helpful discussions.

\end{document}